\documentclass[a4paper,11pt,leqno]{article}
\makeatletter
\usepackage{amsfonts,delarray,amssymb,amsmath,amsthm,a4,a4wide}
\usepackage{xcolor,color,epsfig,graphics}

\newtheorem{theo}{Theorem}

\newtheorem{conjecture}{Conjecture}
\newtheorem{defi}{Definition}

\def\Xint#1{\mathchoice
   {\XXint\displaystyle\textstyle{#1}}%
   {\XXint\textstyle\scriptstyle{#1}}%
   {\XXint\scriptstyle\scriptscriptstyle{#1}}%
   {\XXint\scriptscriptstyle\scriptscriptstyle{#1}}%
   \!\int}
\def\XXint#1#2#3{{\setbox0=\hbox{$#1{#2#3}{\int}$}
     \vcenter{\hbox{$#2#3$}}\kern-.5\wd0}}
\def\dashint{\Xint-}

\def\({\left(}
\def\){\right)}
\def\1{\mathbf{1}}

\def\bo{\partial \Omega}

\def\curl{{\rm curl\,}}

\def\dt0{{{\frac{d}{dt}}_{|t=0}}}

\def\ep{\varepsilon}

\def\hal{\frac{1}{2}}

\def\hciii{{H_{c_3}}}
\def\hcii{{H_{c_2}}}
\def\hci{{H_{c_1}}}

\def\he{{h_{\rm ex}}}

\def\indic{\mathbf{1}}
\def\indic{\mathbf{1}}

\def\io{\int_{\Omega}}

\def\l{{\ell}}

\def\lep{{|\mathrm{log }\ \ep|}}

\def\llep{\log\ \lep}

\def\l|{\left|}

\def\mc{\mathbb{C}}

\def\mr{\mathbb{R}}

\def\mz{\mathbb{Z}}
\def\nab{\nabla}

\def\np{\nab^{\perp}}

\def\om{\Omega}

\def\p{\partial}
\def\Q{\mathcal{Q}}

\def\ro{\rho}
\def\r|{\right|}

\def\sm{\setminus}

\def\T{{\mathbb{T}}}

\def\vp{\varphi}

\def\z{\zeta}
\def\Z{Z_n^\beta}
\def\Q{\mathbb{P}_n^\beta}
\def\E{\Sigma}
\def\W{\mathcal{W}}

\def\R{{\mr}}

\begin{document}

\title{Ginzburg-Landau vortices, Coulomb Gases, and  Renormalized  Energies}

\author{Sylvia Serfaty}

\maketitle

\begin{abstract} This is a review about a series of results on vortices in the Ginzburg-Landau model of superconductivity on the one hand, and point  patterns in Coulomb gases on the other hand, as well as the connections between the two topics.
\end{abstract}

{\bf keywords:} Ginzburg-Landau equations, superconductivity, vortices, Coulomb Gas, one-component plasma, jellium, Renormalized energy.

Most of this paper describes joint work  with Etienne Sandier, which has naturally led  from the study of the Ginzburg-Landau equations of superconductivity -- a rather involved system of PDE -- to that of  a well-known  statistical mechanics system: namely the classical Coulomb gas. 
We will review results in each area and explain the similarity in the mathematics involved.

\section{The Ginzburg-Landau model of superconductivity}

A type-II superconductor, cooled down below its  critical temperature experiences the circulation of ``superconducting currents" without resistance, and  has a particular response in  the presence of an applied magnetic field. Above a certain value of the external field called the {\it first critical field}, {\it vortices} appear.  When the field is large enough, the experiments  (dating from the 60's) show that they  arrange themselves in (often) perfect triangular lattices, cf. {\tt http://www.fys.uio.no/super/vortex/} or Fig. \ref{fig0} below.

\begin{figure}[h]\begin{center}
\includegraphics[height=5cm]{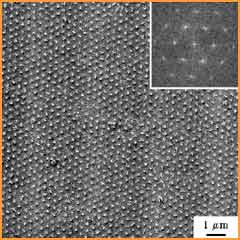}
\includegraphics[height=5cm]{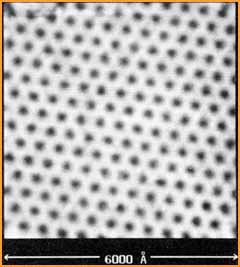} \\
\caption{Abrikosov lattices}\label{fig0}\end{center}\end{figure}

These are named {\it Abrikosov lattices} after the physicist Abrikosov who had predicted, from the Ginzburg-Landau model, that periodic arrays of vortices  should appear \cite{abri}.  These vortices repel each other like  Coulomb  charges would, while being confined inside  the sample by the applied  magnetic field. Their triangular lattice arrangement is the result of these two opposing effects.

\subsection{The model}
The Ginzburg-Landau model was introduced on phenomenological grounds by Landau and Ginzburg in the 50's \cite{gl} ; 
after some nondimensionalizing procedure, in a two-dimensional domain $\om$ it takes  the form 
\begin{equation}\label{gl}
 GL_\ep (\psi,A)
 = 
 \frac{1}{2} \int_{\om} |(\nab-iA)\psi |^2 + |\nab \times A-h_{ex}|^2+ \frac{1}{2\ep^2} \(1-|\psi|^2\)^2
 \end{equation}
 
 Here $\psi$ is a complex-valued function, called ``order parameter" and indicating the local state of the sample: $|\psi |^2$ is the density of ``Cooper pairs" of
superconducting electrons. With our normalization $|\psi|\le 1$, and where
$|\psi|\simeq 1$ the material is in the superconducting phase, while
where $|\psi|= 0$, it is in the normal phase (i.e. behaves like a
normal conductor), the two phases being able to coexist in the
sample.

The vector field $A$ is the gauge field or vector potential of the magnetic field. The
magnetic field in the sample is deduced by $h=\nab \times A=\curl A = \p_1
A_2-\p_2 A_1$, it is thus a real-valued function in $\om$.

Finally, the parameter $\ep$ is  a material constant, it is the inverse of the
``Ginzburg-Landau parameter" usually denoted $\kappa$. It is also  the ratio   between the ``coherence length" usually denoted $\xi$ (roughly the vortex-core size) and the  ``penetration length" of the magnetic field usually denoted $\lambda$.      We are interested in the regime of small $\ep$, corresponding to
high-$\kappa$ (or extreme type-II) superconductors. The limit $\ep
\to 0$ or $\kappa \to \infty$ that we will consider is also called
the London limit.

This is a $\mathbb{U}(1)$-gauge theory and the  functional (as well as all the physically meaningful quantities) is invariant under the gauge-change
\begin{equation}
\label{1.51} \left\{
\begin{array}{ll}
\psi \mapsto \psi e^{i \Phi}\\
A\mapsto A + \nab \Phi
\end{array}
\right.
\end{equation}
where $\Phi$ is a smooth enough function.

The stationary
states of the system are the critical points of $GL_\ep$, or the
solutions of the Ginzburg-Landau equations :
\begin{equation*}\leqno{\mathrm{(GL)}}
\left\{
 \begin{aligned} -(\nabla_A)^2 \psi= \frac{1}{\ep^2} \psi(1-|\psi|^2) &
  \quad \text{in $\om$}\\
 -\nabla^\perp h = \langle i\psi,\nabla_A \psi \rangle  &\quad \text{in $\om$}\\
 h = \he &\quad \text{on $\bo$}\\
 \nabla_A \psi \cdot \nu  = 0  & \quad \text{on $\partial\om$,}\\
 \end{aligned}
 \right.
 \end{equation*}
where $\np $ denotes the operator $(-\p_2,\p_1)$, $\nab_A=\nab -i A$, and $\nu$ is the
outer unit normal to $\bo$.
Here appears the {\it superconducting current}, a real valued  vector field given by
$j=\langle i\psi,\nab_A \psi\rangle$ where $\langle.,.\rangle $ denotes
the scalar-product in $\mc$ identified with $\mr^2$. It may also
be written as
$$\frac i2\(\psi \overline{\nab_A \psi}  - \bar \psi \nab_A \psi\),$$ where
the bar denotes the complex conjugation.
For further  details on the model, we refer to \medskip \cite{gl,dg,t,fh,livre}.

The Ginzburg-Landau   model  has led to a large amount of theoretical physics literature --- probably  most relevant to us is the book by De Gennes \cite{dg}. However,  a precise mathematical proof of the phase transition at the first critical field, and of the emergence of the Abrikosov lattice as the ground state for the arrangement of the vortices was still missing. 

In the 90's, researchers  coming from nonlinear analysis and PDEs became interested in the model (precursors were Berger, Rubinstein, Schatzman, Chapman, Du, Baumann, Phillips... cf. e.g. \cite{ch,dugu} for reviews),  with the notable contribution of Bethuel-Brezis-H\'elein \cite{bbh}  who introduced   systematic  tools and  asymptotic estimates to study vortices, but in the simplified Ginzburg-Landau equation not containing the magnetic gauge, and allowing only for a fixed number of vortices. This was then adapted to the model with gauge but with a different boundary condition  by Bethuel and Rivi\`ere \cite{br,br2}. It was however not clear that this approach could work to treat the case of the full magnetic model when the number of vortices gets unbounded as $\ep \to 0$. It is only with the works of Sandier \cite{sa} and Jerrard \cite{jerr} that tools capable of handling this started to be developed.
Relying on these tools and expanding  them,  in  a series of works later revisited in a book \cite{livre}, we  analyzed the full model and obtained the proof of the phase transition, and the computation of the asymptotics of the first critical field in the limit   $\ep \to 0$.
We characterized the optimal number and distribution of the vortices and derived in particular a ``mean-field regime" limiting distribution for the vortices, which  will be described just below.
Note that this analysis and  the tools developed to understand the vortices  have proven useful to study vortices in rotating superfluids like Bose-Einstein condensates (cf. e.g.  \cite{roug} and references therein), a problem which has a large  similarity  with Ginzburg-Landau from the mathematical perspective, and of current interest for experiments. 

\subsection{Critical fields and vortices}
What are vortices?  A vortex
is an object centered at an isolated zero of $\psi$, around which the
phase of $\psi$ has a nonzero winding number, called the {\it degree
of the vortex}. So it is also a small defect of normal phase in the superconducting phase, surrounded by a loop of superconducting current. 
 When $\ep$ is small, it is clear from
\eqref{gl} that  any discrepancy between $|\psi|$ and $1$ is strongly penalized,   and a scaling
argument hints that $|\psi|$ is different from $1$ only  in regions of
characteristic size $\ep$.
 A typical vortex centered at a point $x_0$
behaves  like $ \psi= \rho e^{i\,\varphi} $ with $\ro= f(\frac{|x-x_0|}{\ep}) $ where $f(0)=0$ and $f$ tends to $1$
as $r \to + \infty$, i.e. its characteristic core size is $\ep$,
and
$$\frac{1}{2\pi} \int_{\p B(x_0,R\ep)} \frac{\p \varphi}{\p \tau} = d \in \mz$$ is
an integer, called the {\it degree of the vortex}. For example
$\vp= d \theta$ where $\theta $ is the polar angle centered at
$x_0$ yields a vortex of degree $d$ at $x_0$.

There are three main critical values of $\he$ or {\it critical
fields} $\hci$, $\hcii$, and $\hciii$, for which phase-transitions
occur. 
\begin{itemize}
\item For $\he<\hci$ there are no vortices and  the energy minimizer is  the superconducting state $(\psi \equiv 1, A\equiv 0)$. (This is a true solution if $\he=0$, and a solution close to this one (i.e. with $|\psi|\simeq 1$ everywhere) persists if $\he $ is not too large.)
It is said that the superconductor ``expels" the applied magnetic field, this is the ``Meissner effect", and the corresponding solution is called the Meissner solution. 
\item  For $\he= \hci$, which is of the order of $\lep$ as $\ep \to 0$,   the first vortice(s) appear.
\item For $\hci<\he<\hcii$ the superconductor is in the ``mixed phase" i.e. 
there are vortices, surrounded by superconducting phase where $|\psi|\simeq 1$.
 The higher $\he>\hci$, the more vortices
there are. The vortices  repel each other so they tend to arrange in
these triangular Abrikosov lattices in order to minimize their
repulsion.
\item  For $\he= \hcii\sim \frac{1}{\ep^2} $, the vortices are so densely packed that they
 overlap each other, and  a second phase transition
 occurs, after which $|\psi|\sim 0$ inside the sample, i.e. all
 superconductivity in the bulk of the sample is lost.
\item For $\hcii<\he<\hciii$
 superconductivity
persists only  near the boundary, this is called {\it surface
superconductivity}.  More details and the mathematical study of this transition are found in \cite{fh} and references therein.
\item 
 For $\he> \hciii=O(\frac{1}{\ep^2}) $ (defined
in decreasing fields), the sample is completely in the normal
phase, corresponding to the ``normal" solution $\psi\equiv 0, h\equiv \he$ of (GL). See \cite{gp} for a proof. 
\end{itemize}

\subsection{Formal correspondence}
Given a family of configurations $(\psi_\ep, A_\ep)$ it 
turns out to be convenient to express the energy in terms of the induced magnetic field $h_\ep(x) = \nab \times A_\ep(x)$. Taking the curl of the second relation in (GL) we obtain
$$-\Delta h_\ep+ h_\ep= \curl \langle i\psi_\ep,\nabla_{A_\ep} \psi_\ep \rangle + h_\ep\simeq \curl \nab \vp_\ep$$
where we approximate $|\psi_\ep|$ by $1$ and where $\vp_\ep$ denotes the phase of $\psi_\ep$.
One formally has that $\curl \nab \vp_\ep= 2\pi \sum_i d_i \delta_{a_i}$ where $\{a_i\}_i$ is the collection of  zeroes of $\psi$, i.e. the vortex centers (really depending on $\ep$), and $d_i\in \mz$ are their topological degrees. This is not exact, however it can be given some rigorous meaning in the asymptotics $\ep \to 0$. We may rewrite this equation in a more correct manner
\begin{equation}\label{london}
\left\{\begin{array}{ll}
-\Delta h_\ep + h_\ep \simeq 2\pi \sum_i d_i \delta_{a_i}^{(\ep)} & \text{in} \ \om\\
h=\he & \text{on} \ \bo,\end{array}\right.\end{equation}
where the exact right-hand side in \eqref{london} is a sum of quantized  charges, or Dirac masses,  which  should be thought of as somehow  smeared out at a scale of order  $\ep$.
This relation is called in the physics literature the {\it London equation} (it is usually written with true Dirac masses, but this only holds approximately). It indicates  how the magnetic field penetrates in the sample through the vortices.

Some computations  (with the help of all the mathematical machinery developed to describe vortices) lead eventually to the conclusion that everything happens as if the Ginzburg-Landau energy $GL_\ep$ of a configuration were equal to  
\begin{multline}
\label{gep}
GL_\ep (\psi_\ep, A_\ep) \simeq \hal \io |\nab h_\ep|^2 + |h_\ep - \he|^2 \\
=\hal
\iint_{\om \times \om } G_{\om} (x,y) \Big(2\pi \sum_i d_i \delta_{a_i}^{(\ep)} - \he\Big)(x) \Big( 2\pi \sum_i d_i \delta_{a_i}^{(\ep)}-\he\Big) (y),\end{multline}
where $G_\om$ is a type of Green kernel, solution to 
\begin{equation}\label{gom}
\begin{cases}- \Delta G_\om + G_\om= \delta_y   & \text{in } \ \om\\
G_\om= 0 & \text{on} \ \bo,\end{cases}\end{equation} 
and $h_\ep $ solves \eqref{london}.
With this way of writing, and in view of the logarithmic nature of $G_\om$, one recognizes essentially a pairwise Coulomb interaction of positive charges in a constant negative background ($-\he$), which is what leads to the analogy with the Coulomb gas described later. 
There remains to understand for which value of $\he$ vortices become favorable, and with which distribution. To really understand that, the effect of the ``smearing out" of the Dirac charges needs to be more carefully accounted for. Instead of each vortex having any infinite cost in \eqref{gep} (which would be the case with true Diracs) the real cost of each vortex can be evaluated as being $\sim \pi d_i^2  \lep$ per vortex (roughly the equivalent of the cost generated by a Dirac mass  smeared out at the scale $\ep$). 
We may thus evaluate \eqref{gep} as 
\begin{equation}
\label{geptrue}
\frac{GL_\ep(\psi_\ep, A_\ep)}{\he^2} \simeq \frac{\lep}{\he} \frac{\pi \sum_i d_i^2}{\he}+\frac{\he^2 }{2}\io |\nab h|^2 + |h-1|^2 \end{equation}
where $h= \lim_{\ep \to 0} \frac{h_\ep}{\he}$.

Optimizing over the degrees $d_i$'s allows to see that the degrees $d_i=1$ are the only favorable ones. In view of \eqref{london}, assuming this is true we can then rewrite $\pi \sum_i d_i^2 $ as $\hal\io |-\Delta h_\ep + h_\ep|$. Passing to the limit $\ep \to 0$, and assuming $\frac{\he }{\lep } \to \lambda$ as $\ep \to 0$, we find 
 that the mean-field limit  energy arising from \eqref{gep} is 
\begin{multline}\label{mf} \frac{GL_\ep}{\he^2} \simeq_{\ep \to 0}  \mathcal{E}^{MF}_{\lambda} (h) = \frac{1}{2\lambda} \io |-\Delta h+h|+ \hal \io |\nab h|^2 + |h-1|^2 \\ = \frac{1}{2\lambda} \io |\mu| + \hal \iint_{\om \times \om} G_\om(x,y) d(\mu-1)(x)\, d(\mu-1)(y)\end{multline}
 where  $h$ is here related to the limiting  ``vorticity" (or vortex density) $\mu:= \lim_{\ep \to 0} \frac{1}{\he} 2\pi\sum_i d_i\delta_{a_i}$  by $-\Delta h+h= \mu$ in $\om $ with $h=1$ on $\bo$, simply by taking the limit of \eqref{london}.

In \eqref{mf} the first contribution to the energy corresponds the total self-interaction of the vortices in \eqref{gep}, while the second one is the cross-interaction of the vortices and the vortices and the equivalent background charge (which is really the result of the confinement effect of the applied magnetic field).

We have the following rigorous statement.
\begin{theo}[\cite{ss3}, \cite{livre} Chap. 7] \label{theo14}
Assume $\he \sim \lambda \lep$ as $\ep \to 0$,  where $\lambda>0$ is a constant
independent of $\ep$. If  $(\psi_\ep, A_\ep) $ minimizes $GL_\ep$, then as $\ep \to 0$
$$\frac{2\pi \sum_i d_i \delta_{a_i}}{\he} \to \mu_*\quad \text{in the weak sense of measures} $$ where $\mu_*$ is the unique
minimizer of $\mathcal{E}^{MF}_\lambda$. Moreover \begin{equation}\label{mf2}
\min GL_\ep = \he^2(\min \mathcal{E}^{MF}_\lambda +o(1))  \quad \text{as}  \ \ep \to 0.\end{equation}
\end{theo}

Minimizing $\mathcal{E}^{MF}_\lambda$ leads to  a standard variational problem called an ``obstacle problem". The corresponding optimal distribution of vorticity  is uniform of density $1-1/(2\lambda)$ on a subdomain $\omega_\lambda$ of $\om$ depending only on $\lambda$.

\begin{figure}[h]
\begin{center}
\includegraphics[height=5cm]{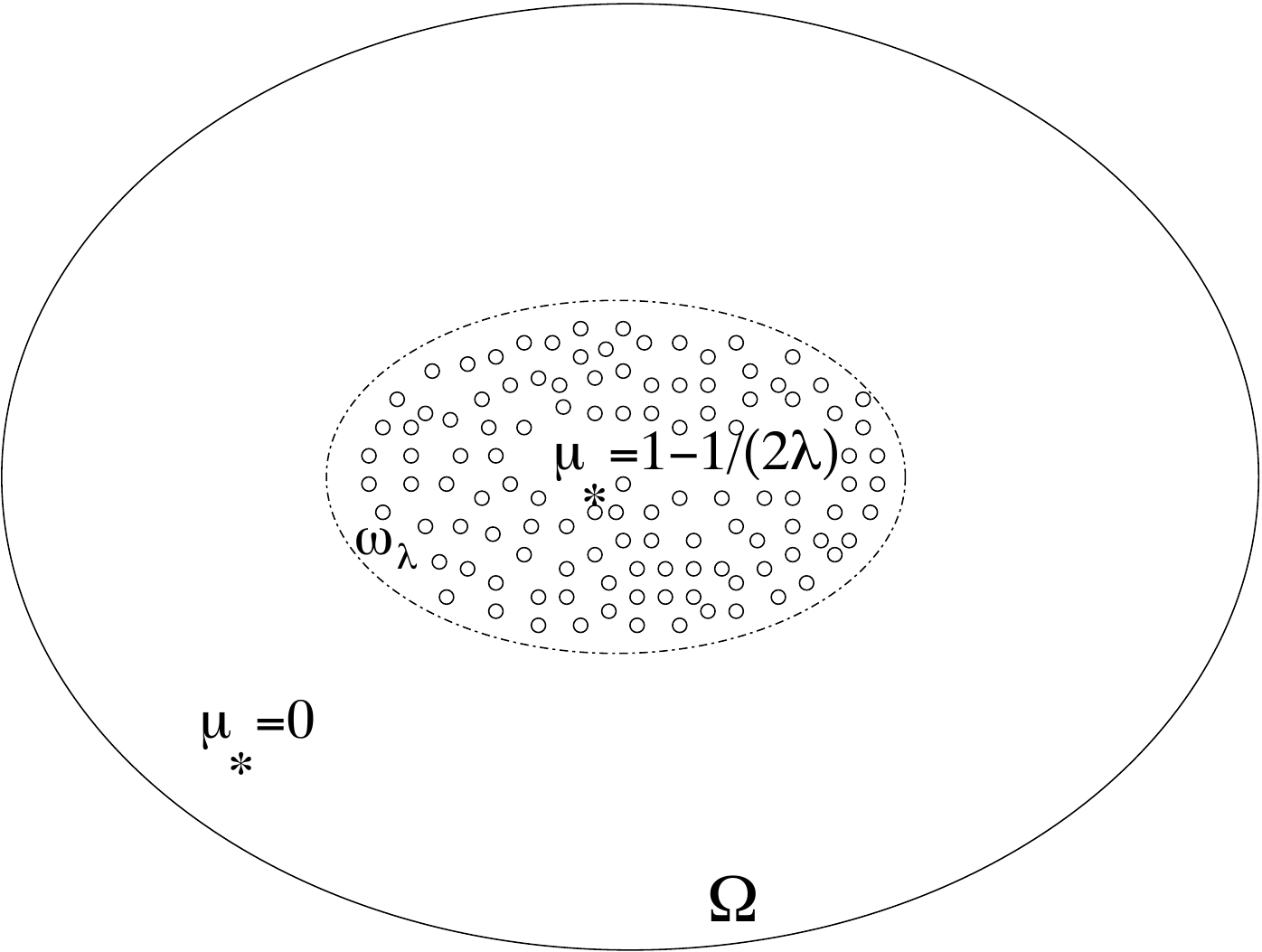}\end{center}
\caption{Optimal density of vortices according to the obstacle
problem.}\label{fig3}
\end{figure}

An easy analysis of this obstacle
problem yields the following (cf. also Fig. \ref{fig3}):
\begin{enumerate}
\item $\omega_\lambda= \varnothing$ (hence $\mu_*=0$) if and only
if $\lambda <\lambda_\om$, where $\lambda_\om$ is given by
\begin{equation}\label{k0}\lambda_\om = (2\max |h_0-1|)^{-1}\end{equation} 
for $h_0$ the solution to
\begin{equation}\label{h0}\left\{\begin{array}{ll}
- \Delta h_0+h_0=0& \text{in} \ \om\\
h_0= 1& \text{on} \ \bo.\end{array}\right.\end{equation}
 \item For $\lambda >\lambda_\om$,
 the measure of $\omega_\lambda$ is nonzero, so the limiting
vortex density $\mu_*\neq 0$. Moreover, as $\lambda $ increases
(i.e. as $\he$ does), the set $\omega_\lambda $ increases. When $\lambda =
+\infty$ (this
corresponds to the case $\he \gg \lep$), 
$\omega_\lambda $ becomes $\om$ and $\mu_*=\indic_\om$.\end{enumerate}

Since $\hci$ corresponds to the applied field for which minimizers start to have vortices, this leads us to expecting that 
\begin{equation}\label{hc1}
\boxed{ \hci\sim \lambda_\om \lep \quad \text{as} \ \ep \to 0}
\end{equation}
where $\lambda_\om$ depends only on $\om$ via \eqref{k0}--\eqref{h0}.

In fact this is true, because we were able to show that below this value $\hci$, not only the average vortex density $\mu_*$ is $0$, but there are really no vortices.
To see this, a more refined asymptotic expansion of $GL_\ep$ than \eqref{mf} is needed.
It suffices instead to note that in the regime when a zero or  small number of vortices is expected, the solution to \eqref{london} is well approximated by $\he h_0$ where $h_0$ solves \eqref{h0}, and then to split the true $h_\ep$ as $\he h_0 + h_{1,\ep}$ where $h_{1,\ep}$ is a remainder term, and expand the energy $GL_\ep$ in terms of this splitting.

Let us state the result we obtain when looking this way more carefully at the regime $\he\sim \lambda_\om \lep$ and analyzing individual vortices. For simplicity, we assume that the function $h_0$ achieves a unique minimum at a point $p \in \om$ (this is satisfied for example if $\om $ is convex)  and that its Hessian at that point, $Q$, is nondegenerate.
\begin{theo}[\cite{s1,s2}, \cite{livre} Chap. 12]\label{thpetitn}
There exists an increasing sequence of values 
$$H_n = \lambda_\om\lep +
(n-1)\lambda_\om \log \frac{\lep}{n} + \text{constant order terms}$$
such that if $\he\le \lambda_\om \lep + O(\llep)$ and $\he \in (H_n, H_{n+1})
$, then global
minimizers of $GL_\ep$ have exactly $n$ vortices of degree 1, at points
$a_i^\ep \to {p}$ as $\ep \to 0$, and the $\tilde{a_i^\ep} :=
\sqrt{\frac{\he}{n}} (a_i^\ep-p) $ converge as $\ep \to 0$ to a
minimizer of
\begin{equation}\label{wn0}w_n(x_1,\cdots, x_n) = - \sum_{i
\neq j} \log |x_i-x_j|+ n \sum_{i=1}^n Q(x_i).\end{equation}
\end{theo}
We find here the precise value of $\hci$ for which the first vortex appears in the minimizers, and then a sequence of ``critical fields" for which a second, a third, etc.. vortices appear in minimizers, together with a characterization of their optimal locations, governed  by an explicit interaction energy $w_n$.

\subsection{The next order study}

Theorem \ref{theo14} above proved that above $\hci$, for $\lambda>\lambda_\om$,  the number of vortices is proportional to $\he$ and they are  uniformly distributed in a subregion of the domain, but  it is still  far from explaining the optimality of the Abrikosov lattice.  To (begin to) explain it, one needs to look at the next order in the energy asymptotics \eqref{mf}, and at the blown-up of \eqref{london} at the inverse of the intervortex distance scale, which here is simply $\sqrt{\he}$. For simplicity, let us reduce to the case $\lambda=1$ (or $\he \gg \lep$) where the limiting optimal measure is $\mu_*=  \indic_{\om}$ and the limiting $h\equiv 1$.
 
Once the blow-up by $\sqrt{\he}$ is performed and the limit $\ep \to 0$ is taken, \eqref{london} becomes  
\begin{equation}\label{londonbu}
-\Delta H +1 = 2\pi \sum_a \delta_a\quad \text{in} \ \mr^2\end{equation} where the limiting blown-up points $a$ form an infinite configuration in the plane, and these are now true Diracs (one may in fact reduce to the case where all degrees are equal to $+1$, other situations being energetically too costly). 

One may  recognize here essentially a {\it  jellium} of infinite size, and $E=\nab H$ the electric field generated by the  points (its rotated vector field $j=- E^\perp$ corresponds to the superconducting current in superconductivity).
The  jellium model was first introduced by Wigner, and it means an infinite   set of  point charges with identical charges with Coulomb interaction, {\it screened} by a uniform neutralizing background, here the density $-1$.  It is also called a {\it one-component \medskip plasma}.

It then remains first  to identify and define a limiting interaction energy for this ``jellium," and second to derive it from $GL_\ep$. The energy, that will be denoted $W$, arises as a next order correction term in the expansion of $\min GL_\ep$ beyond the order $\he^2$ term identified by \eqref{mf}-\eqref{mf2}.

Isolating efficiently the next order terms in $GL_\ep$ relies on another ``splitting" of the energy. Instead of expanding $h_\ep$ near $\he h_0$ as in the case with few vortices, one should expand around $\he h_\lambda$ where $h_\lambda $ is the minimizer of $\mathcal{E}^{MF}_\lambda$, i.e. the  solution  to the obstacle problem above.
Splitting $h_\ep= \he h_\lambda + h_{1,\ep}$ where $h_{1,\ep}$ is seen as a remainder, turns out to exactly isolate the leading order contribution $\he^2 \min \mathcal{E}^{MF}_\lambda$ from an explicit lower order term.

\subsubsection{The renormalized energy: definition and properties}
We next turn to discussing the effective interaction energy between the blown-up points. As we said, it should be a total Coulomb interaction between the points (seen as discrete positive point charges) and the fixed constant negative background ``charge". Of course defining the total Coulomb interaction of such a system is delicate  because several difficulties arise:  first,  the infinite number of charges and the lack of local charge neutrality, which lead us to defining the energy as a thermodynamic limit; second the need to remove the infinite self-interaction created by each point charge, now that we are dealing with true Diracs.

Let us now define the interaction energy $W$.
 Let $m>0$ be a  given positive number (corresponding to the density of points). 
We say a vector field $E$ belongs to the class $\mathcal{A}_m$ if 
\begin{multline}\label{eqj}
 E= \nab H\quad  -\Delta H = 2\pi (\nu - m)   \quad\mbox{for some} \  \nu=  \sum_{p \in\Lambda} \delta_{p}, \quad \mbox{where } \ \Lambda \ \mbox{is a discrete set in \ }\mr^2.
\end{multline} 
As said above,
the vector-field $E$ physically corresponds, in the electrostatic analogy, to the electric field generated by the point charges, and $-E^\perp$ to a superconducting current  in the superconductivity context. 

Note that 
 $H$ has a logarithmic singularity near each $a$, and thus $|\nab H|^2$ is not integrable; however, when removing small balls  of radius $\eta$ around each $a$, adding back $\pi \log \eta$, and letting $\eta \to 0$,  this singularity can be ``resolved".

\begin{defi}
We define the {\it renormalized energy} $W$ for $E \in \mathcal{A}_m$  by
\begin{equation}\label{defw}
W(E):= \limsup_{R \to \infty}
\frac{W(E, \chi_{B_R})}{|B_R|},
\end{equation}
where $\chi_{B_R}$ is any cutoff function supported in $B_R$ with $\chi_{B_R} =1$ in $B_{R-1}$ and $|\nab\chi_{B_R}| \leq C$, and $W(E, \chi)$ is defined by
\begin{equation}\label{WW}
W(E,  \chi):=
\lim_{{\eta\to 0}}  \int_{\mr^2\sm\cup_{i=1}^n B(x_i,  \eta)}{\chi} |E|^2
+ \pi{ ( \log \eta )}\sum_i {\chi}(x_i).
\end{equation}\end{defi}

The name is given by analogy with the ``renormalized energy" introduced in \cite{bbh}
 as the effective interaction energy of a finite number of point vortices. Renormalized refers here to the way the energy is computed by  substracting off the infinite contribution corresponding to the self interaction of each charge or vortex.

In the particular case where the configuration of points $\Lambda$ has some periodicity, i.e. if it can be seen as $n$ points $a_1, \cdots,a_n $ living on a torus $\T$ of appropriate size, then $W$ can be expressed much more simply as a function of the points only:
\begin{equation}\label{ara}W(a_1, \cdots, a_n)= \frac{\pi}{|\T|} \sum_{j \neq k} G(a_j- a_k) + \pi \lim_{x\to 0} \( G(x)+ \log |x|\),\end{equation}
where $G$ is the  Green's function of the torus (i.e. solving $- \Delta G=\delta_0- 1/|\T|$). The Green function of the torus can itself be expressed explicitely in terms of some Eisenstein series and the Dedekind Eta function. The definition  \eqref{defw} thus allows to generalize such a formula to any infinite system, without any periodicity assumption.

The question of central interest to us is that of understanding the minimum and minimizers of $W$. Here are a few remarks.
\begin{enumerate}
\item The value of $W$ doesn't really depend on the cutoff functions satisfying the assumption.
\item $W$ is unchanged by a compact perturbation of the points.
\item One can reduce by scaling to studying $W$ over the class $\mathcal{A}_1$.
\item It can be proven that minimizers of $W$ over $\mathcal{A}_1$ exist (and the minimum is finite).
\item It can be proven that the minimum of $W$ is equal to the limit as $N \to \infty$  of the minimum of $W$ over configurations of points which are $N\times N$ periodic.
\end{enumerate}

We do not know the value of $\min_{\mathcal{A}_1} W$, however we can identify the minimum of $W$ over a restricted class: that of points on a perfect lattice (of volume $1$).

\begin{theo}[\cite{ss}]
The minimum of  $ W$ over perfect lattice configurations (of density $1$) is achieved  uniquely, modulo rotations, by the triangular lattice.
\end{theo}
By triangular lattice, we mean the lattice $\mathbb{Z}+ \mathbb{Z}e^{i\pi/3}$, properly scaled.

The proof of this theorem uses the explicit formula for $W$ in the periodic case  in terms of Eisenstein series
mentioned above.
By transformations using modular functions or by direct computations,
 minimizing $W$ becomes equivalent to minimizing the Epstein
 zeta function  $\zeta(s)= \sum_{p \in \Lambda} \frac{1}{|p|^s}
 $, $s>2$,  over lattices. Results from number theory  in the 60's to 80's (due to
 Cassels, Rankin, Ennola, Diananda, Montgomery, cf. \cite{montgomery} and references therein) say
 that this is minimized by the triangular lattice.

In view of the experiments showing Abrikosov lattices in superconductors,  it is  then natural to formulate the
  \begin{conjecture} \label{conj1}
The ``Abrikosov" triangular lattice is a global minimizer of $W$.
\end{conjecture}
Observe that
this question  belongs to the more general family of crystallization problems. A typical such question is,  given a potential $V$ in any dimension, to determine the point positions that minimize
$$\sum_{i\neq j} V(x_i-x_j)$$ (with some kind of boundary condition), or rather
$$\lim_{R\to \infty}\frac{1}{|B_R|}\sum_{i\neq j, x_i, x_j \in B_R} V(x_i-x_j),$$
and to determine whether the minimizing configurations are perfect lattices.  Such questions are fundamental in order to understand
the  crystalline structure of matter.
 They also arise in  the arrangement of   Fekete points, ``Smale's problem" on the sphere,  or the  ``Cohn-Kumar conjecture"...
One should immediately stress that there are very few positive results in that direction in the literature  (in fact it is very rare to have a proof  that the solution to some  minimization problem is periodic). Some exceptions include the two-dimensional sphere packing  problem, for which Radin \cite{radin} showed that the minimizer is the triangular lattice, and an extension of this by Theil \cite{theil} for a class of very  short range Lennard-Jones potentials.
Here the corresponding potential is rather logarithmic, hence long-range, and these techniques  do not apply.
 The question could also  be rephrased as that  of finding
$$\min`` \|\sum_i \delta_{x_i}-1\|_{(H^{1})^*}"  $$
where the quantity is put between brackets to recall that $\delta_{x_i}$ does not really belong to the dual of the Sobolev space  $H^{1}$ but rather has to be computed in the renormalized way that defines $W$. A closely related problem: to find
 $$ \min \|\sum_i \delta_{x_i} -1\|_{Lip^*}$$  turns out to be  much easier. It is  shown in \cite{bpt} with a relatively short proof that again the triangular lattice achieves the minimum.
 
With S. Rota Nodari, in \cite{rns}, we investigated further the structure of minimizers of $W$ (or rather, a suitable modification of it) and we were able to prove that the energy density and the points were uniformly distributed at any scale $\gg 1$, in good agreement with (but of course much weaker than!) the conjecture of periodicity of the minimizers.

Even though the minimization of $W$ is only conjectural, it is natural to  view it as (or expect it to be)  a quantitative ``measure of disorder" of a configuration of points in the plane.
    In this spirit, in  \cite{bs}
    we use $W$ to quantify and compute explicitly  the disorder of some classic random point configurations in the plane and on the real line.

\subsubsection{Next order result for $GL_\ep$}
We can now state the main next order result on $GL_\ep$.

\begin{theo}[\cite{ss}]\label{th11}  Consider minimizers $(\psi_\ep,A_\ep)$ of the Ginzburg-Landau in the regime $\lambda_\om \lep \le \he \ll \frac{1}{\ep^2}$.
After blow-up  at scale $\sqrt{\he}$ around a randomly chosen point in $\omega_\lambda$, the  $\nab  h_\ep $
converge as $\ep \to 0$ to vector fields  in the plane which, almost surely, minimize $W$.
Moreover, $ \min GL_\ep  $ can be computed up to $o(\he)$ (i.e. up to an error $o(1)$ per vortex):
 $$\min GL_\ep=  \he^2 \min \mathcal{E}^{MF}_\lambda + (1-\frac{1}{2\lambda}) \he |\omega_\lambda| \min_{\mathcal{A}_1} W+o(\he)\quad \text{as} \ \ep \to 0.$$ 
\end{theo}

Thus, our study reduces the problem to understanding the minimization of $W$. If the last step of proving Conjecture \ref{conj1}  was accomplished, this would completely justify the emergence of the Abrikosov lattice in superconductors. 

\subsection{Other results}
We also investigated the structure of solutions to (GL) which are not necessarily minimizing,  in other words critical points of \eqref{gl}. Let us list here the main results:
\begin{itemize}
\item In \cite{s2} and \cite[Chap 12]{livre}, we prove the existence of branches of local minimizers of \eqref{gl} (hence stable solutions) of similar type as the solutions in Theorem \ref{thpetitn}
which have arbitrary bounded numbers of vortices all of degree $+1$ and exist for wide ranges of the parameter $\he$, and the locations of the vortices in these solutions are also characterized.
 In other words, for a given $\he$, there may exist an infinite number of stable solutions with vortices, indexed by the number of vortices. Only one specific value of the number of vortices is optimal, depending on the value of $\he$, as in Theorem \ref{thpetitn}. 
\item Similarly, there also exist multiple branches of  locally minimizing solutions of (GL) with (rather arbitrary) unbounded numbers of vortices. This is proven in \cite{cs}. These vortices arrange themselves according to a uniform density over a set again determined by an obstacle problem, and at the microscopic level, they tend to minimize $W$.
\item If one considers a general solution to (GL) with not too large energy, then one can characterize the possible distributions of the vortices, depending on whether their number is bounded or unbounded as $\ep \to 0$. The characterization says roughly that the total force (generated by the other vortices and by the external field) felt by a vortex has to be zero. In particular it implies that if there is a large number of vortices converging to a certain regular density, that density must be constant on its support.
This is proven in \cite{ss4} and \cite[Chapter 13]{livre}.\end{itemize}
There are also many results on the dynamics of such vortices under heat or Schr\"odinger versions of (GL). Most of them concern a finite number of vortices, for example \cite{lin,js,bos1,bos2,bos3,spirn,spirn2,s7}, with the exception of \cite{ks}.

The analysis of the three-dimensional version of the Ginzburg-Landau model is of course more delicate than that of the two-dimensional one, because of the more complicated geometry of the vortices, which are lines instead of points (the first result attacking this question was \cite{ri1}).
This explains why it has taken more time to get analogous results. The best to date is the three-dimensional equivalent of Theorem \ref{theo14} by Baldo-Jerrard-Orlandi-Soner \cite{bjos1,bjos2}, see also references therein and \cite{kach}.

\section{The 2D Coulomb gas}

The connection with the jellium is what prompted us to examine in \cite{ma2d} the consequences that our study could have for   the 2D classical Coulomb gas.  More precisely, let us consider a 2D Coulomb gas of $n$ particles  $x_i \in \mr^2$ in a confining potential $V$ (growing sufficiently fast at infinity), and let us  take the mean-field scaling of interaction where the Hamiltonian is given by
  \begin{equation}\label{wn}
  H_n(x_1, \cdots, x_n)= -  \sum_{i \neq j} \log |x_i-x_j|+ n \sum_{i=1}^n  V(x_i).\end{equation} (At least if the potential $V$ is homogeneous, one may always reduce to this case by scaling.)
  Note that ground states of this energy  are also  called ``weighted Fekete sets", they arise in interpolation, cf. \cite{st}, and are interesting in their own right.
  
\begin{figure}[h]
\begin{center}
\includegraphics[height=6cm]{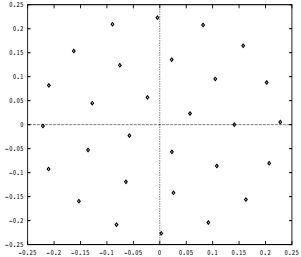}
\caption{Numerical minimization of $H_n$ by Gueron-Shafrir \cite{gs}, $n=29$}
\label{fig4}

\end{center}\end{figure}

For $V(x)=|x|^2$, some numerical simulations, see Fig. \ref{fig4}, give the shapes of minimizers of $H_n$, which is then also a particular case of $w_n$ that appeared in \eqref{wn0}.

The Gibbs measure for the same  mean-field Coulomb gas at temperature $1/\beta$ is 
\begin{equation}\label{cg}
d\Q(x_1, \cdots, x_n)=   \frac{1}{\Z} e^{-\beta H_n (x_1, \cdots, x_n)}dx_1\, \cdots dx_n
\end{equation}
where $\Z$ is the associated  partition function, i.e. a normalization factor that makes $d\Q$ a probability measure. The particular case of $\beta=2$ and $V(x)= |x|^2$ corresponds to the law of  the eigenvalues for random matrices with iid normal entries, the {\it Ginibre ensemble}.  The connection between Coulomb gases and random matrices was first pointed out  by Wigner \cite{wigner} and Dyson \cite{dyson}.  For general background and references, we refer to \cite{forrester}.
The same situation but with $x_i$ belonging to the real line instead of the plane is also of importance for random matrices, the corresponding law is that of what is often called a ``$\beta$-ensemble".

  Among interacting particle  systems, Coulomb gases have always been considered as particularly  interesting but delicate, due to the long range nature of the interactions (which is particularly true in 1 and 2 dimensions).   The case of one-dimensional Coulomb gases can be solved more explicitly \cite{len,kunz,bl,am}, and  crystallisation at zero temperature is established.
In dimension 2, many studies rely on  a rather ``algebraic approach"  with exact computations (e.g. \cite{janco2}), or require a finite system or a condition ensuring local charge balance \cite{aj,sm}.  Our approach is strictly energy-based and this way valid for any temperature  and general potentials $V$.

  \subsection{Analysis of ground states}
  A rather  simple analysis of \eqref{wn}, analogous to \eqref{mf},  leads to the result that minimizers of $H_n$ satisfy $\frac{\sum_{i=1}^n \delta_{x_i}}{n}\to \mu_0$ where $\mu_0$ minimizes the  following mean-field limit for $H_n/n^2$ as $n \to \infty$:
  \begin{equation}
\label{I}  \mathcal{F}(\mu)= \int_{\mr^2\times \mr^2} - \log |x-y|\, d\mu(x) \, d\mu(y) + \int_{\mr^2} V(x)\, d\mu(x)\end{equation}
defined for $\mu$ a probability measure. The  unique mean-field minimizer, which is also called the {\it equilibrium measure} in potential theory is a probability $\mu_0$ (just as the minimizer of $\mathcal{E}^{MF}_\lambda$  for Ginzburg-Landau, it can also be viewed as the solution of an obstacle problem). Its support, that we will denote $\E$, is compact (and assumed to have a nice boundary). For example if $V(x)=|x|^2$,  it is a multiple of  the characteristic function of a ball (the {\it circle law} for the Ginibre ensemble in the context of random matrices), and this is analogous to  Theorem \ref{theo14} and the  obstacle problem distribution found for Ginzburg-Landau.  Deriving this mean-field limit is significantly easier than for Ginzburg-Landau, due to the discrete nature of the starting energy, and the fact that all charges are $+1$ (as opposed to the vortex degrees, which can be any integer).

The connection with the Ginzburg-Landau situation is made by defining analogously the potential generated by the charge configuration using  the mean-field density $\mu_0$ as a neutralizing  background, this yields the following equation playing the role of the  analogue to \eqref{london}:
\begin{equation*}
 h_n= - 2\pi\Delta^{-1} \Big(\sum_{i=1}^n \delta_{x_i} - n \mu_0\Big) \quad \text{in} \ \mr^2.\end{equation*}
The next step is again  to express this
in the blown-up coordinates at scale $\sqrt{n}$ (analogous to the $\sqrt{\he}$ scale previously) around $x_0$,  $x'=  \sqrt{n} (x -x_0) $, via 
$h_n'$  the solution to
\begin{equation} \label{split2}
 h_n'(x') = -2\pi \Delta^{-1} \Big(\sum_{i=1}^n \delta_{x_i'} -  \mu_0 (x_0+   \frac{ x'}{ \sqrt{n}} ) \Big).\end{equation}When taking $n\to \infty$, the limit equation to \eqref{split2} is
 \begin{equation}\label{Hlim}-\Delta h= 2\pi \Big( \sum_{a} \delta_{a} - \mu_0(x_0)\Big)\quad \text{in} \ \mr^2\end{equation}
 analogue of \eqref{londonbu},
 corresponding to another infinite jellium with uniform neutralizing background $\mu_0(x_0)$.
 
Expanding the energy to  next order is done via a suitable splitting, by analogy with Ginzburg-Landau. In fact in this setting the splitting procedure is quite simple: it suffices to write  $\nu_n:= \sum_{i=1}^n \delta_{x_i}$ as $n \mu_0 + (\nu_n- n \mu_0)$. Noting that
\begin{equation*}H_n(x_1, \cdots, x_n)= \iint_{\triangle^c} - \log |x-y|\, d\nu_n(x) \, d\nu_n(y)
+\int V(x) \, d{\nu_n(x)}\end{equation*}
where $\triangle $ denotes the diagonal, inserting  the indicated splitting of $\nu_n$, 
 we eventually  find the exact decomposition
\begin{equation}\label{split1}
H_n(x_1, \cdots, x_n)= n^2 \mathcal{F}(\mu_0)-\frac{ n}{2} \log n + \frac{1}{\pi} W(\nab   h_n', \indic_{\mr^2} )+2 n \sum_{i=1}^n  \z (x_i), \end{equation}
where $W(E,  \chi)$ is as in \eqref{WW}.
The function $\zeta$ in \eqref{split2} is explicit and determined only by $V$, it is like an effective potential and  plays no other role than confining the particles to $\E=Supp (\mu_0)$ (it is zero there, and positive elsewhere), so there remains to understand the limit $n \to \infty$ of
$W(\nab h_n', \indic_{\mr^2})$ corresponding to  \eqref{Hlim}. One of our main results below is that this term is of order $n$.  
An important advantage of this formulation is that it transforms, via \eqref{split1}, the sum of pairwise interactions into an extensive quantity in space \eqref{WW}, which allows for localizing (via a screening procedure), cutting and pasting, etc...

Let us now state the next order result playing the role of Theorem \ref{th11}.

\begin{theo}[\cite{ma2d}]\label{th1} \label{th51}
Let  $(x_1,\dots, x_n)\in (\mr^2)^n$.
 Then
 \begin{equation}\label{binf}
  \liminf_{n \to \infty}\frac{1}{n} \(H_n(x_1, \dots, x_n) - n^2 \mathcal{F} (\mu_0)   + \frac{n}{2}\log n \) \ge  \frac{|\E|}{\pi}\int W(E) \, dP(x,E),\end{equation}
where  $P$ - a.e. $(x,E) \in \mathcal{A}_{\mu_0(x)}$,  and  $P $ is a probability,  limit of the   push-forward of  $\frac{1}{|\E|}dx_{|\E}$ (the normalized Lebesgue measure on $\E$)  by
$$x\mapsto \(x, E_{n}(\sqrt{n}x+\cdot) \), \qquad E_n:=\nab h_n'.$$
This lower bound is sharp; thus  for minimizers of $H_n$,  for   $P$-a.e. $(x,E)$, $E$ minimizes $W$ over $\mathcal{A}_{\mu_0(x)}$ and
\begin{multline}\label{minhn}
\lim_{n\to \infty}\frac{1}{n}\(\min H_n - n^2 \mathcal{F} (\mu_0)   + \frac{n}{2}\log n \)= \frac{|\E|}{\pi} \int\Big( \min_{E \in \mathcal{A}_{\mu_0(x)} } W\Big)\, dP(x,E)\\
= \frac{1}{\pi}\int_\E \min_{\mathcal{A}_{\mu_0(x)} }W :=\alpha_0.\end{multline} \end{theo}

This result contains a sharp lower bound valid for any configuration, and not just for minimizers. The lower bound is a sort of average (expressed via the probability $P$) of the renormalized energy $W$, with respect to all the possible blow-up centers $x_0\in \E$. A rephrasing is that minimizers of $H_n$ provide configurations of points in the plane whose associated ``electric fields" $E$ minimize, after blow-up and taking the limit $n\to \infty$, the renormalized energy, $P-$a.e., i.e. (heuristically) for almost every blow-up center. $W$ is a sort  of next-order limiting energy for $H_n$ (or next order $\Gamma$-limit, in the language of $\Gamma$-convergence).
 It is the term that sees the difference between different microscopic patterns of points, beyond the macroscopic averaged behavior $\mu_0$.

If again  the conjecture on the minimizers of $W$ was established, this would  prove that points in  zero temperature Coulomb  gases should form a crystal in the shape of an Abrikosov triangular \medskip lattice. 

 The result  of Theorem \ref{th51} can be improved when one looks directly at minimizers (or ground states) of $H_n$ instead of general configurations. With S. Rota Nodari, we obtained the following 

\begin{theo}[\cite{rns}]\label{thsimona} Let $(x_1,\dots,x_n)$ be a minimizer of $H_n$. Let  $x_i'$, $\nab h_n'$, $\E$, $\mu_0$, be as above and $\E'= \sqrt{n} \E$, $\mu_0'(x)= \mu_0(x/\sqrt{n})$. The following holds:
\begin{enumerate}
	\item\label{itempointth3} for all $i\in [1,n]$, $x_i\in \Sigma$;
	\item\label{itemequidistth3} there exist $\beta \in (0,1)$, $\bar c>0$, $C>0$ (depending only on $\mu_0 $),  such that for every $\ell\ge \bar c$ and $a\in \E'$ such that $d (K_{\ell }(a), \p \E') \ge  n^{\beta/2}$, we have
	\begin{equation}\label{137}
	\limsup_{n\to \infty}\frac{1}{\ell^2}\left| W(\nab h_n',\chi_{K_{\ell}(a)} )-\int_{K_\ell(a)}\Big(\min_{\mathcal A_{\mu_0'(x)}}W\Big)\,dx\right|\le o(1) \quad \text{as } \ell \to+\infty,
	\end{equation}
	where $\chi_{K_{\ell(a)}}$ is any cutoff function supported in $K_\ell(a) $ and equal to $1$ in $K_{\ell-1} (a)$; 
	and
	\begin{equation}\label{eqnumpointsgas}
\limsup_{n\to \infty}	\left| \# (K_\ell(a) \cap \{ x_i'\}) -\int_{K_{\ell}(a)}m'_0(x)\,dx\right|\le C\ell.
	\end{equation}
\end{enumerate}
\end{theo}
This says that for minimizers, the configurations seen after blow-up around {\it any} point sufficiently inside $\E$ (and not just a.e. point) tend to minimize $W$ and their points follow the distribution $\mu_0$ even at the microscopic scale.

This result is to be compared with previous results of Ameur - Ortega-Cerd\`a \cite{aoc} where, using a very  different method based on ``Beurling-Landau densities", they prove \eqref{eqnumpointsgas},  with a larger possible error $o(\ell^2)$  but for distances to $\E'$ which can be smaller (their paper does not however contain the \medskip connection with $W$).

When \eqref{wn} is considered for $x_i\in \mr$ instead of $\mr^2$, then it is the Hamiltonian of what is called a ``log gas". The same corresponding result are proven in \cite{ss5}, together with the definition of an appropriate one-dimensional version of $W$, for which the minimum is this time shown to be achieved by the lattice configuration (or ``clock distribution") $\mz$.

\subsection{Method of the proofs}\label{sec2.1}
The proof of the above Theorems \ref{th11} and \ref{th51} follows the idea of $\Gamma$-convergence (see e.g. \cite{braides,dalmaso}) i.e. relies on obtaining general (i.e. ansatz-free) lower bounds, together with matching upper bounds obtained via explicit constructions.

There are really two scales in our energies: a macroscopic scale (that of the support of $\mu_*$ or $\mu_0$), and the scale of the distance between the points (or vortices) which is much smaller.  We know how to obtain lower bounds for the energy at the microscopic scale, but it is not clear in our case how to ``glue" these estimates together.  For that purpose we introduced in \cite{ss} a new general method for  obtaining lower bounds on two-scale energies.
  A probability measure approach allows to integrate the local estimates  via the use of the ergodic theorem (an idea suggested by S. R. S.  Varadhan).
That abstract  method applies well to positive (or bounded below) energy densities, but  those associated to $W(E,\chi)$  are not,   due to one of the main difficulties mentioned above: the lack of local charge neutrality. To remedy this
we start by modifying the energy density to make it bounded below, using  sharp energy lower bounds  by improved ``ball construction" methods (\`a la
        Jerrard / Sandier).
        
        The method is the same for both cases but in the case of the Coulomb gas, it is complicated  by the (slow) variation of the macroscopic density $\mu_0$.
         For Ginzburg-Landau, the  situation is on the other hand made more difficult  by the presence of vortices of arbitrary signs and degrees.

Finally, let us mention that in \cite{gms1,gms2} we carry out a similar analysis for the ``Ohta-Kawasaki" model of diblock-copolymers, where the interacting objects are this time ``droplets" that can have more complicated geometries and nonquantized charges (their charge is really their volume), and derive the same next order limiting energy $W$.

\subsection{The case with temperature}
Understanding the asymptotics of $H_n$ via Theorem \ref{th51} (and not only of ground states) naturally  allows to deduce  information on  finite temperature states.
First, inserting the lower bound on $H_n$ found in Theorem \ref{th51} into \eqref{cg} directly yields an upper bound on $\Z$.  Conversely, using an explicitly constructed test-configuration meant to approximate minimizers of $H_n$ up to $o(n)$, and showing that a similar upper bound holds in a sufficiently large phase-space  neighborhood of that configuration, allows to deduce a lower bound for $\Z$. The lower bound and the upper bound will coincide as $\beta \to \infty $ only. The main statement is 
\begin{theo}[\cite{ma2d}] We have, for $n$ large enough,
\label{th52} 
 \begin{equation*}
n\beta f_1(\beta) \leq \log \Z - \Big( -\beta n^2 {\cal F} (\mu_0) + \frac{\beta n}{2} \log n \Big) \leq n \beta f_2(\beta),
\end{equation*}
where $f_1(\beta)$ and $f_2(\beta)$ are independent of $n$, bounded, and
\begin{equation*}
\lim_{\beta \to \infty} f_1(\beta) = \lim_{\beta \to \infty} f_2(\beta) = 
 \frac{1}{\pi}   \int  \min_{\mathcal{A}_{\mu_0(x)}} W\, dx =\alpha_0.
\end{equation*} \end{theo}
Only the term in $n^2$ of this expansion was  previously known, for such a general situation of general $\beta $ and $V$. This is in contrast to the one-dimensional log gas case where $\Z$ is known for $V$ quadratic and all $\beta$ by Selberg integrals, and for more general $V$'s as well.
Also, the result relates the computation of $\Z$ to that of the unknown constant $\min W$, so to prove  Conjecture \ref{conj1} it would suffice in principle  to know how to \medskip  compute $\Z$ for a 2D Coulomb gas!

The final result is a large deviations type result. First, let us 
 recall the best previously  known result which is a  result of ``large deviations from the circle law":
\begin{theo}[Petz-Hiai \cite{hiaipetz}, Ben Arous-Zeitouni \cite{bz}, ]\label{th5}
$\Q$ satisfies a large deviations principle with good rate function $\mathcal{F}(\cdot)$ and speed $n^{-2}$: 
for all $A \subset \{ \mbox{probability measures} \}$,
\begin{equation*}
-\inf_{\mu \in A^\circ} \widetilde {\cal F }(\mu) \leq \liminf_{n\to \infty} \frac{1}{n^2} \log \Q(A) \leq \limsup_{n\to\infty} \frac{1}{n^2} \log \Q(A) \leq -\inf_{\mu\in \bar A} \widetilde {\cal F} (\mu),
\end{equation*}
where $\widetilde{\mathcal{F} }= \mathcal{F} - \min \mathcal{ F}$.
\end{theo} 
This thus says that  the probability of an event $A$ is exponentially small if $\cal F>\min \cal F $ in $A$, i.e. if one deviates from the circle law $\mu_0$:
$$\Q(A) \le e^{-n^2 \inf_{\bar{A}}\({\cal F} - {\cal F }(\mu_0)\)}.$$

\begin{figure}\begin{center}
\includegraphics[height=6cm]{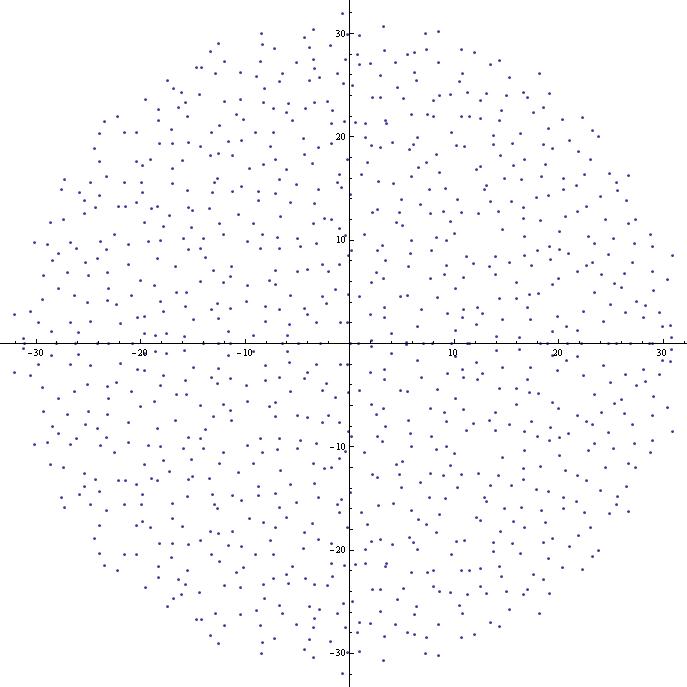} \\
\caption{Eigenvalues of 1000-by-1000 matrix with i.i.d Gaussian entries ($\beta=2$)} \label{fig5}
(from Benedek Valk\'o's webpage)\end{center}\end{figure}

One may wonder whether the same is true at next order, i.e. whether the arrangements of points after blow up follow the next order optimum of $H_n$, i.e. minimize $W$.
Figure \ref{fig5}, corresponding to the Ginibre case of $\beta=2 $ and $V(x)=|x|^2$   indicates that this should not  be the case since  the points do not arrange themselves according to triangular lattices.

We may then wonder how to quantify the order  or rigidity of the configurations according to the temperature. The following result provides some answer, and an improvement at next order on Theorem \ref{th5}:
\begin{theo}[\cite{ma2d}]   \label{th53}
Let $A_n \subset (\mr^2)^n$. Then, for every $\ep>0$ there exists $C_\ep$ (bounded when $\ep $ is bounded away from $0$) such that 
\begin{equation*}
\limsup_{n\to \infty} \frac1n \log \Q(A_n) \leq -\beta \Big(  \frac{|\E|}{\pi} \inf_{P\in A}\int W(E) dP(x,E) - \alpha_0- \ep - \frac {C_\ep}{\beta} \Big),
\end{equation*}
and $A$ is the set of probability measures which are limits (in the weak sense) of blow-ups at rate $\sqrt{n}$ around a point $x$ of the electric fields $\nab h_n$ associated to $ \sum_{i=1}^n \delta_{{x_i}}$ with $(x_i) \in A_n$.
\end{theo}

Modulo again the conjecture on $\min W$, this proves  crystallization  as the temperature goes to $0$:
after blowing up around a point $x$ in the support of $\mu_0$, at the scale of $\sqrt{n}$, as $\beta\to \infty$  we  see (almost surely) a configuration which minimizes $W$. Indeed, $\alpha_0$ is the minimum value that $\inf_{P\in A}\int W(E) dP(x,E)$ can possibly  take, and is achieved if and only if $W(E)=\min_{\mathcal{A}_{\mu_0(x)} } W $ for $P$-a.e. $(x,E)$.  

For $\beta$ finite, the result says that the  average of $W$ lies below a fixed threshhold (say $\alpha_0+1+ \frac{C}{\beta}$), except with very small probability.

To our knowledge, this is the first time Coulomb gases are rigorously
connected to triangular lattices (again modulo the solution to the conjecture on the minimum of $W$), in agreement with predictions in the  literature (cf. \cite{aj} and references therein).

\section{Higher dimensional Coulomb Gases}
With Nicolas Rougerie, in \cite{rs}, we extended the results for the Coulomb gas presented above to arbitrary  higher dimension, considering this time 
$$H_n(x_1,\dots, x_n)= \sum_{i\neq j} g(x_i-x_j)+ n \sum_{i=1}^n  V(x_i)$$
with $x_i \in \mr^d$ and the kernel $g$ is the Coulomb kernel $-\log |x|$ in dimension $2$ and $|x|^{2-d}$ in dimension $d\ge 3$.
The mean-field limit energy  is defined in the same way by 
$$\mathcal{F}(\mu)= \iint_{\mr^d\times \mr^d} g(x-y) \, d\mu(x) \, d\mu(y)+ \int_{\mr^d} V(x)\, d\mu(x).$$

Turning to higher dimension  required a new approach and a new definition of $W$, the previous one being very tied to the two-dimensional ``ball construction method" as alluded to in Section \ref{sec2.1}.
The new approach is based on a different way of renormalizing, or substracting off the infinite ``self-interaction" energy of each point: we replace 
 point charges by smeared-out charges, as in ``Onsager's lemma". More precisely,  we pick some arbitrary fixed {\it radial} nonnegative function $\ro$, supported in $B(0,1)$ and with integral $1$, and  for any point $x$ and $\eta>0$ we introduce   the smeared charge 
\begin{equation*}
\delta_x^{(\eta)}= \frac{1}{\eta^d}\ro \left(\frac{\cdot}{\eta}\right) *  \delta_x.
\end{equation*}
Newton's theorem asserts that the Coulomb potentials generated by the smeared charge $\delta_x^{(\eta)}$ and the point charge $\delta_x$ coincide outside of $B(x,\eta)$. A consequence of this is that if we define   instead of 
\begin{equation}\label{28}
h_n'(x') = - c_d \Delta^{-1}  \( \sum_{i=1}^n \delta_{x_i'} - \mu_0 (x_0 + x' n^{-1/d})\) 
\end{equation}
as in \eqref{split2}, the potential
\begin{equation}\label{}
h_{n,\eta}'(x')  = - c_d \Delta^{-1} \( \sum_{i=1}^n \delta_{x_i'}^{(\eta)} - \mu_0 (x_0 + x' n^{-1/d})\)\end{equation} 
then $h_{n}' $ and $h_{n,\eta}'$ coincide outside of the $B(x_i', \eta)$. Moreover, they differ by 
$\sum_i  f_\eta(x-x_i')$ where $f_\eta$ is  a fixed function equal to $c_d \Delta^{-1}(\delta_0^{(\eta)}-\delta_0)$, vanishing outside $B(0,\eta)$.
(Here the constant $c_d$  is the constant such that $-\Delta g= c_d \delta_0$, depending only on dimension).

By keeping these smeared out charges, we are led at the limit to 
solutions  to 
\begin{equation}-\Delta h_\eta = c_d \Big( \sum_a \delta_a^{(\eta)}- m\Big), \quad m = \mu_0(x_0)\end{equation}
which are in bijection with the functions $h$  solving the same equation with $\eta=0$, via adding or subtracting $\sum_a f_\eta(x-a)$.


For any fixed $\eta>0$ one may then define the electrostatic energy per unit volume of the infinite jellium with smeared charges as
\begin{equation}\label{eq:Weta pre}
 \limsup_{R\to \infty} \dashint_{K_R}  |\nab h_\eta|^2 := \limsup_{R\to \infty} |K_R| ^{-1} \int_{K_R}  |\nab h_\eta|^2 
\end{equation}
where $h_\eta$ is as in the above definition and $K_R$ denotes the cube $[-R,R]^d$. This energy is now well-defined for $\eta>0$ and blows up as $\eta \to 0$, since it includes the self-energy of each smeared charge in the collection. One should then ``renormalize" \eqref{eq:Weta pre} by removing the self-energy of each smeared charge before taking the limit $\eta \to 0$. The leading order  energy of a smeared charge is $\kappa_d g(\eta)$ where $\kappa_d$ is a constant depending on dimension and on the choice of the smoothing function $\ro$.
We are then led to the definition  
\begin{defi}[\cite{rs}]
The renormalized energy $\W$ is defined over the class $\mathcal{A}_m$ by
\begin{align}\label{weta}
\W(\nab h) = \liminf_{\eta\to 0}\W_\eta(\nab h) \nonumber =\liminf_{\eta\to 0}\left(\limsup_{R\to \infty} \dashint_{K_R}   |\nab h_\eta|^2 - m(\kappa_d g(\eta)+\gamma_2 \indic_{d=2})\right),
\end{align} where $\kappa_d$  and $\gamma_2$ are constants depending only on the choice of $\ro$.
\end{defi}
It is also proven in \cite{rs} that $\W$ achieves its minimum (for each given $m$), which in dimension $2$ coincides with that of $W$.  It is also natural to expect that the minimum of $\W$ may be achieved by crystalline configurations (like the FCC lattice in three dimensions) but this is a completely open question.

We are able to show that a similar splitting relation as \eqref{split1} holds, which makes appear $\W_\eta$ instead of $W$. It is however only an inequality, but equality is retrieved as one lets $\eta \to 0$.  This allows to let $n\to \infty$ and obtain lower bounds via the same ``probabilistic method"  mentioned in Section \ref{sec2.1}, for fixed $\eta $. At the end we let $\eta $ tend to $0$, to retrieve similar results as Theorems  \ref{th52}-\ref{th53}.

The following  gives the analogue to Theorem \ref{th52} but this time expressed in terms of the free energy $F_{n,\beta}= - \frac{2}{\beta} \log \Z$.

\begin{theo}[\cite{rs}]\label{thm:partition}\mbox{}\\
Let us define \begin{equation}\label{eq:gamma d}
 \xi_d := \begin{cases}\displaystyle
             \frac{1}{c_d} \left(\min_{\mathcal{A}_1}  \W \right) \int_{\R   ^d} \mu_0 ^{2-2/d} \quad \mbox{if } d\geq 3\\
            
             \displaystyle \frac{1}{2\pi} \min_{\mathcal{A}_1}  \W - \hal \int_{\R   ^2} \mu_0 \log \mu_0  \quad \mbox{if } d=2.
            \end{cases}
\end{equation}
 If $d\geq 3$ and $\beta \geq c n ^{2/d - 1}$ for some $c>0$ we have, for $n$ large enough, for any $\ep>0$
\begin{equation}\label{eq:partition low T}
\left| F_{n,\beta} - n ^2 \mathcal{F}  (\mu_0)  - n ^{2-2/d}   \xi_d   \right| \leq \ep  n^{2-2/d} + C_\ep\frac{n}{\beta},
\end{equation}  respectively if $d=2$ and $\beta \geq c (\log n) ^{-1} $, $c>0$, for $n$ large enough, for any $\ep>0$, 
\begin{equation}\label{eq:partition low T 2d}
\left| F_{n,\beta} - n ^2 \mathcal{F} (\mu_0) + \frac{n}{2}\log n - n \xi_2 \right| \leq  \ep n + C_\ep \frac{n}{\beta},
\end{equation} where $C_\ep$ depends only on $\ep$, $V$ and  $d$ and is bounded when $\ep$ is bounded away from $0$.
\end{theo}

Taking in particular formally $\beta =\infty$ leads to $F_{n, \infty}= \min H_n$ and allows to retrieve the next order    expansion  of $\min H_n$.

An analogue of Theorem \ref{th53} is also given:

\begin{theo}[\cite{rs}]\label{th4}
For any $n>0$ let $A_n\subset (\mr^d)^n$. Assume that $\beta \ge c n ^{2/d-1}$ for some constant $c>0$. Then for every $\ep>0$, there exists a constant $C_\ep$ depending only on $\ep$, $V$, and bounded when $\ep $ is bounded away from $0$,   such that we have
\begin{equation}
\label{ldr}
\limsup_{n\to \infty} \frac{ \log \Q(A_n)}{n ^{2-2/d}} \le -\frac{\beta}{2} \left(\frac{|\E|}{c_d} \inf_{P\in A_\infty}\int \W(E) dP(x,E) -( \xi_d +\ep) - C_\ep \lim_{n\to \infty} \frac{n ^{2/d-1}} \beta \right),
\end{equation}
where $A$ is the set of probability measures which are limits (in the weak sense) of $\nab h_n'(x+ \cdot)$  associated to the $(x_i)\in A_n$ via \eqref{28}.

\end{theo}

These results  show that the expected regime for crystallization (maybe surprisingly) depends on the dimension and is  the regime  $\beta \gg n^{2/d-1}$. In that regime,  the Gibbs measure essentially concentrates on minimizers of $\W$, which as before would show crystallization if one knew that such minimizers have to be crystalline.

For a self-contained presentation of these topics, one can also refer to \medskip  \cite{ln}.

Other than proving that specific crystalline configurations achieve the minimum of $W$ or $\W$, the main result missing in these studies is to establish a complete Large Deviations Principle at next order for the Coulomb gas, and identifying the right rate function which should involve $W$, but not only. This would prove at the same time the existence of a complete next order ``thermodynamic limit" (for leading order results see \cite{liebnarn} and references therein.

Let us finish by pointing out that  the quantum case (of the Coulomb gas)  is quite different and studied in \cite{lnss}:  the next order  term is of order $n$ and identified to be the ground state of the ``Bogoliubov Hamiltonian."

\noindent
{\sc S. Serfaty\\
UPMC Univ  Paris 06, UMR 7598 Laboratoire Jacques-Louis Lions,\\
 Paris, F-75005 France ;\\
 CNRS, UMR 7598 LJLL, Paris, F-75005 France \\
 \&  Courant Institute, New York University\\
251 Mercer st, NY NY 10012, USA.\\}
{\tt E-mail: serfaty@ann.jussieu.fr}

\end{document}